\documentclass[twocolumn,showpacs,preprintnumbers,amsmath,amssymb]{revtex4}
\makeatletter
\def\@dotsep{4.5}
\makeatother

\def\tf  {t_{\rm fold}}
\def\H   {{\cal H}_{\rm eff}^{(m)}}
\def\Hn   {{\cal H}_{\rm eff}^{(n)}}
\def\ds {\displaystyle}
\def\erfc{{\rm erfc}}
\def\d {{\rm d}}
\def\dr{\partial}
\def\tijm{T_{ij}^-}
\def\tijp{T_{ij}^+}

\usepackage{graphicx}% Include figure files
\usepackage{dcolumn}% Align table columns on decimal point
\usepackage{bm}% bold m

\begin{document}

\preprint{}

\title{
Folding Kinetics of Proteins and Cold Denaturation.
}
%\shorttitle{Title} %Insert here a short version of the title if it exceeds 70 characters

\author{Olivier Collet}

\affiliation{LPM, Nancy-Universit\'{e}, CNRS, \\
Boulevard des Aiguillettes BP 239, F-54506 Vandoeuvre-l\`{e}s-Nancy}

\date{\today}

\begin{abstract}
Folding kinetics of a lattice model of protein is studied.
It uses the
Random Energy Model for the intrachain couplings and a temperature
dependent free energy of solvation derived from a realistic hydration
model of apolar solutes.
The folding times are computed using Monte Carlo simulations
in the region of the phase diagram 
where the chain occurs in the native structure.
These folding times are roughly equals for the temperatures of cold and warm denaturation 
for a large range of solvent quality.
Between these temperatures, the folding times reach maxima and thus, 
at low temperatures, the kinetics of the chain always speeds up as 
the temperature is decreased.
The study of the conformational space as function of the temperature 
permits to elucidate this phenomenon. 
At low temperature, it shows that the activation barriers of the system decrease faster than
the temperature as the temperature is decreased.
At high temperature, the rate of the barriers over the temperature decreases as the 
temperature is increased because the height of the barrier is almost constant.
\end{abstract}

%{Folding dynamics}
%{Protein-solvent interactions}
%{Generic models (lattice, HP, etc.)}

\maketitle

Proteins are very long molecular chains built with given sequences
of amino-acids. Under biological solvent conditions, a protein 
occurs in a unique, native, compact form \cite{Anfinsen1961} 
and an important goal
of theoretical physics is to understand how a chain finds
its native structure in a reasonable biological time.

Lattice models, in which the amino acids of the chain are located on
the vertices of a two or three-dimensional lattice, are widely used to study
protein folding.  
In the Random Energy Model (REM)\cite{Bryngelson1987,Shakhnovich1990a,Dinner1994,Shakhnovich1994,Gutin1995b}, 
the couplings between monomers
are chosen at random in a Gaussian distribution centered on a negative value.
It leads to an energy spectrum where a few (all compact) conformations lye in the bottom discrete part of the spectrum while the large majority of the conformations belongs to a
quasi-continuous  top part. This spectrum may well
mimics that of proteins. However, although REM  
 explains some features of the proteins, 
it is independent on the temperature and it fails to reproduce a
general feature specific to proteins: the cold denaturation\cite{Kumar2006,Pastore2007,Hadi2007,Whitten2007}.

The cold denaturation has been first observed by Privalov \cite{Privalov1989}
for Myoglobin which is in its native form
between a temperature of warm denaturation $T_w$, and a temperature of
cold denaturation $T_c$. 
Above $T_w$ or below $T_c$, the protein is in a denatured  state
where a lot of conformations are relevant. 
These temperatures are very sensitive to the pH of the solvent.

Under physiological conditions, the proteins strongly interact 
with the solvent \cite{Kauzmann1959,Warshel1970,Dill1990,Premilat1997,Collet1996,Frauenfelder2006}
and then any simulation of protein folding
must consider a realistic solvent effect on the chain conformations.
Recently \cite{Collet2001, Collet2005}, the temperature dependence 
of the hydrophobic effect has been 
introduced in the couplings of REM 
using realistic solvent model based on a qualitative study of the energy
spectra of the pure solvent and of the solvent around a monomer \cite{Silverstein1999}. 
As a result, the first phase diagram of protein,
where both warm and cold denaturations occur has been
calculated\cite{Collet2001} showing a 
very good accordance with experimental data.
On the other hand, several works has been published on the subject providing
alternative models which were able to exhibits theoretically the cold
denaturation
\cite{Hansen1999,DeLosRios2000,Roccatano2004,Lopez2008,Patel2008,Dias2008}

In this paper, this protein model is first reminded.
The different contributions to the effective couplings between monomers
are shown as functions of the temperature. This result gives an insight of the
modification of conformational space as the temperature is varied.
Thus, the kinetic properties of this model are studied in the native region. 
Folding times are computed
versus a solvation parameter and the temperature,
using Monte Carlo simulations \cite{Metropolis1953,Chan1994}. 
Moreover, the native conformation, the kinetic trap and the transition state are
determined by a study of the phase space.
Last, the unusual behavior of the folding times 
at low temperature is elucidated by taking into account the
temperature dependence of the free energy of the activation barrier.

\section{Protein Model}
\begin{figure}[htbp]
\centerline{\includegraphics[width=2.8in]{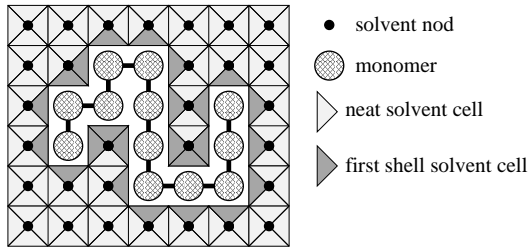}}
\caption{\label{expl}
A 12 monomers chain conformation chosen at random.
}
\end{figure}

The chain is represented by a  string of $N$ beads, (here $N=16$), 
located on a square two-dimensional lattice. 
For a given chain conformation, each empty lattice vertice is considered to be a solvent site.
It is filled up with four solvent cells pointing towards the four
directions (see fig.\ref{expl}). Thus, a solvent cell interacts either with a monomer
or with another solvent cell and its thermodynamic properties are
determined by the type of this interaction.
In such a model, the number of solvent sites and then
the number of solvent cells are constant because the chain length is fixed.
Hence, the volume of the solvent does not
depend on the chain conformation. 
Moreover, here, a unique parameter $B_s$ gives an insight of the solvent quality.
%\vbox{
%}

\section{Effective Hamiltonian}
The system is composed by the chain and a very large number of solvent cells.
It is at equilibrium with a bath at temperature $T$ and
the probability of occurrence of a chain conformation $m$
in interaction with the solvent is~:
$$
P_{\rm eq}^{(m)}(B_s, T) \propto \exp[- \H (B_s,T)/T]$$
The effective Hamiltonian of conformation $m$ takes into account of all
the lattice links and has the following form~:
$$\H = 
\underbrace{\sum_{i > j+1}^N B_{ij} \ \Delta_{ij}^{(m)}}_
{\mbox{intrachain} }
+
\underbrace{\sum_{i=1}^N n_i^{(m)} f_i(T)}_
{\mbox{solvation} }
+
\underbrace{2 
n_s^{(m)} f_s(B_s, T)}_{\mbox{water-water} }   $$
where $B_{ij}$ is the coupling parameter between the monomers $i$ and $j$,
chosen at random in a Gaussian distribution 
centered on 0 with a standard deviation equal to 2 and $\Delta_{ij}^{(m)}$ 
equals 1 if the monomers $i$ and $j$ are 
first neighbors on the lattice and 0 otherwise.
$f_i(T)$ is the free energy of a solvent cell in interaction with monomer $i$ and
$f_s(B_s, T)$ is the free energy of a pure solvent cell,
$n_i^{(m)}$ is the number of links between monomer $i$ and solvent nods
and $n_s^{(m)}$ is the number of solvent vertices bonds.
%%%%%%%%%%%%%%%%%%%%%%%%%%%%%%%%

The expression of the constant number of total lattice links (equals to $K_1$) and 
the fact that each monomer always creates 4 links leads to write the two conservation equations
of the model~:
$$\sum_i \sum_j \frac{1}{2} \Delta_{ij}^{(m)} + \sum_i n_i^{(m)}
+  n_s^{(m)} = K_1$$
$$\sum_j \Delta_{ij}^{(m)} + n_i^{(m)} = 4, 
\  \ \ ({\rm with} \ \ \
\Delta_{i, i+1}^{(m)} = 1)$$

One deduces easily~:
$$\left\{  \begin{array}{l}
n_i^{(m)} = 4 -  \sum_j \Delta_{ij}^{(m)} \\
n_s^{(m)} =  \sum_i \sum_j \frac{1}{2} \Delta_{ij}^{(m)} + K_1 - 4 N
\end{array} \right.$$
Owing to
these relations, and after removing the constant terms, the effective Hamiltonian is rewritten as~:
$$\begin{array}{ll}
\H & = \ds \sum_i \sum_j \left(\frac{1}{2} B_{ij} - f_i +  f_s \right) \Delta_{ij}^{(m)} \\
  & = \ds \sum_i \sum_j \left(\frac{1}{2} B_{ij} - \frac{1}{2}(f_i + f_j) +  f_s \right) \Delta_{ij}^{(m)} \\
  & = \ds \sum_i \sum_{j>i} B_{ij}^{\rm eff}(B_s,T) \Delta_{ij}^{(m)}
\end{array}$$
It takes the form of the usual Hamiltonian without solvent effect 
with effective coupling parameters which now depend on the solvent properties ($B_s$)
and the temperature~:
$$
B_{ij}^{\rm eff}(B_s,T) = B_{ij} - f_i(T) -  f_j(T) +2 f_s(B_s, T) \\
$$

The model used in this work is that introduced in ref.19. The main difference with that used in ref.18 comes from the factor 2 associated to the free energy of a pure solvent cell in the form of the effective couplings. This function guarantees that the total number of solvent cells (i.e. the volume of the solvent) is a constant whatever the chain conformation. In other words, the creation of an link between monomers and j involves the removing of an interaction between residues i in one hand and j in an another hand with a solvent cell each. These two solvent cells becomes pure solvent.

Obviously, the final form of the effective couplings depends on the solvent model used to calculated $f_s$ and $f_i$.

\section{Solvation model}
The solvation free energy calculation are based on results of a study of the hydrophobic effect
undertaken by Dill and coworkers\cite{Silverstein1999}.
They used the Mercedes Benz model of water \cite{Bennaim1970}
and a simple adaptation of the two-states model
of Muller\cite{Muller1990} extended by Lee and Graziano\cite{Lee1996}
to give a physical picture of the hydrophobic effect in terms of two energy
spectra. The first one is associated to the pure solvent and the other one to water molecules in 
interaction with an apolar solute.
%\vbox{
\begin{figure}[htbp]
\centerline{\includegraphics[width=2.8in]{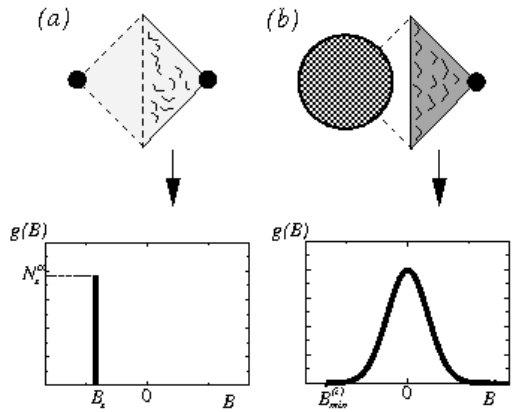}}
\caption{\label{ge}
.(a) Energy spectrum of a cell of pure solvent with a unique $N_s^\alpha$ fold 
degenerated level of energy $B_s$.
(b) Energy spectrum of a cell of  solvent in interaction with a monomer.
$N_s$ energy values are draw at random in a Gaussian distribution centered on 0.
$B_i^{\rm min}$ is specific to each monomer.
}
\end{figure}
In the spectrum associated to pure solvent, the energy gap between the ground 
and the low lying exited states is small.
Here, these two close energy states are gathered in a
unique one of energy $B_s$, $N_s^\alpha$-folds degenerate (see fig. \ref{ge}(a)). 
It comes~:
$$f_s(B_s,T)= B_s - \alpha T \ln N_s \qquad{\rm with} \qquad \alpha < 1$$

In the other spectrum, the energy gap is larger and the exited state is more degenerate.
This picture is reduced further in the spirit of the REM and
the spectrum is spread out (see fig.\ref{ge}(b)). 
For each monomer $i$, $N_s$ values of the solvation energies are draw at random 
in a Gaussian distribution centered on 0 with standard deviation equals 2
and the minimum value of each set of solvation energies,
specific to each monomer is determined and noted $B_i^{\rm min}$.
The free energy of solvent cell in interaction
with the monomer $i$ is then~:
$$f_i(T) = - T \ln \int_{B_i^{\rm min}}^\infty g(B) \exp(-B / T) \d B$$
where $g(B)$ is the density of energy states, i.e., a Gaussian truncated at 
$B_i^{\rm min}$. Thus, one has~:
$$\begin{array}{ll}
f_i(T) 	& =  \ds  -T \ln \int_{B_i^{\rm min}}^\infty \frac{2}{\sqrt{2 \pi}}
\frac{\exp(-B^2/\sigma \sqrt{2})}{\erfc(B_i^{\rm min}/\sigma \sqrt{2}) }
\ \exp\left(-\frac{B}{T}\right) \d B \\
		& = \ds - \frac{\sigma^2}{2 T} - T \ln \left(
N_s \frac
{\erfc\left(\frac{B_i^{\rm min}}{\sigma \sqrt{2}} + \frac{\sigma \sqrt{2}} {2 T }  \right)} 
{\erfc\left(\frac{B_i^{\rm min}}{\sigma \sqrt{2}}  \right)} 
\right)
\end{array}$$
.

%%%%%%%%%%%%%%%%%%%%%%%%%%%%%%%%%%%%%%%%%%%%%%%%%%%%%%%%%%%%%%%%%%%%%

\section{Hydration Results}
The solvation parameters used in this work are $\alpha = 0.5$,  $\sigma=2$ and $N_s = 10^5$. 

As it has already pointed out in a another work\cite{Collet2005}, 
by choosing such solvent parameters, one has~:
$$B_{ij} = B_{ij}^{\rm eff}  -f_i - f_j +2(B_s-\frac{1}{2} T \ln N_s)$$
Then, using $\alpha=0.5$ in the free energy of the pure solvent is equivalent to the 
previous work\cite{Collet2001}
where the solvent parameter has only been divided by 2.

To well understand the kinetic of folding of the chain present later, we first focus
on the different contributions to the effective couplings.
\begin{figure}[htbp]
\vskip0.4cm
\centerline{\includegraphics[width=3.0in]{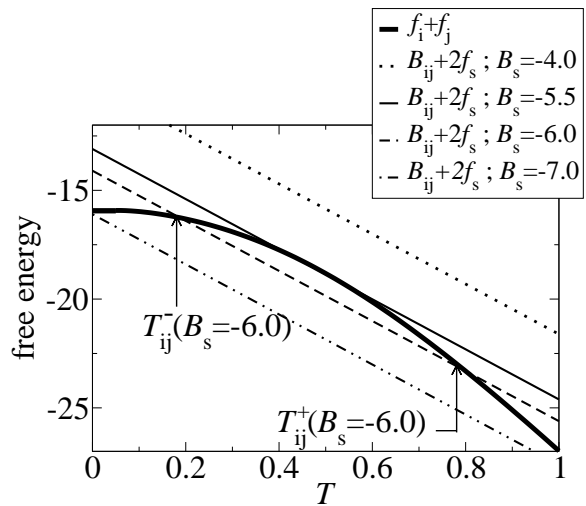}}
\caption{\label{dfifs}
Curves of the different contributions to the effective coupling between the monomers
i=1 and j=4 as function of the temperature for several values of the solvent quality. The two
temperatures for which the coupling vanishes are shown for $B_s=-6.0$.
}
\end{figure}
%}
Figure \ref{dfifs} shows that for large values of $B_s$, the effective couplings are 
repulsive, whatever the temperature, because $B_{ij} + 2 f_s  > f_i + f_j$.
Thus, this condition corresponds to a good solvent in the sense that the monomers
are preferably exposed to the solvent.
For small values of $B_s$, the couplings are attractive at low temperature, modelling
bad solvent condition, and then repulsive at high temperature.
Last, for intermediate values of $B_s$, the curves of $B_{ij} + 2 f_s $ intersect those of
$f_i + f_j$ twice at two temperatures noted $\tijm$ and $\tijp$.
The effective couplings are repulsive for  $T<\tijm$ or $T>\tijp$ and attractive 
for $\tijm < T < \tijp$.

These results show that the free energies of transfer of the residues into water
(here $\delta f_i = f_i - f_s$) present maxima for temperatures, depending
on $B_s$, between $\tijm$ and $\tijp$.
This is in good agreement with studies of the temperature dependence of
the hydrophobic interaction in protein folding. 
As an example, first, Baldwin\cite{Baldwin1986}
showed from calorimetric data, that the transfer of hydrocarbons in water always
exhibits a temperature (denoted as $T_s$) for which the entropy of transfer reaches zero
($\Delta S(T_s) = 0$).
Using the fundamental thermodynamical relation $\Delta S = - \dr \Delta F / \dr T$,
it is clear that the free energy of transfer, $\Delta F$ of the hydrocarbons reaches
a maximum at $T_s$.

\section{Protein Thermodynamics}

It must be noted that, for given solvent quality and temperature,
each intrachain couplings, $B_{ij}$, are different from each other.
The functions $f_i(T)$ are also specific to each residue because the $B_i^{\rm min}$ depend on the
monomers. Thus, some couplings are more attractive than other ones.
As a results, the ground state of the effective Hamiltonian spectra
are non degenerated when the chain is in interaction with a bad quality
solvent ($B_s \ll 0$) .
It corresponds to the native, more maximally compact, conformation
of the sequence.
The native structure (Nat) is determined by a full enumeration of the conformational 
space of the chain in interaction with a bad solvent.
For several values of $B_s$ and $T$, the probability of occurrence of Nat is calculated.
If $P_{\rm eq}^{\rm Nat} > 1/2$, the chain is in the native phase.

%\vbox{
\begin{figure}[htbp]
\centerline{\includegraphics[width=3.0in]{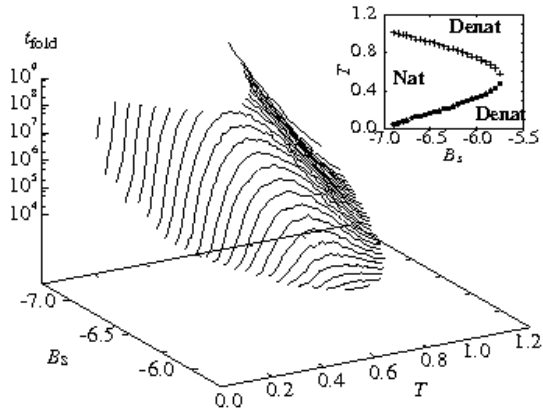}}
\caption{\label{tfold}
Logarithm of the folding time of the chain versus $B_s$ and $T$ 
in the native region for $-7.0 < B_s < -5.75$ and 
$T_c(B_s) < T < T_w(B_s)$.
}
\end{figure}
%}
The inset of fig.\ref{tfold} shows the phase diagram of this sequence.
For $B_s > -5.75$, the chain is always in the denatured 
phase due to the repulsive couplings and then a lot of different structures are relevant. 
For $-7.0 < B_s <-5.75$, the chain occurs in the native state
between $T_c$ and $T_w$ depending on $B_s$.
Below  $T_c$ and above  $T_w$, the chain is denatured.
Obviously the cold denaturation, due to the change in the sign of the effective 
couplings of Nat, occurs for $T_c \approx \tijm$.
Moreover, with the parameters used in this work, one also has $T_w \approx \tijp$.
For temperature smaller than $T_c$, the couplings are mainly repulsive
and then only the numerous chain structures without contact are relevant.
Thus, in the cold phase, the probability of occurrence of the 
Native structure becomes very small. This phase is well denatured because, if
it is a glassy phase, the equilibrium probability of occurrence of the 
native structure would be rather large (and kinetics would be very slow). 
In the cold denatured phase, the set of the extended structures becomes the relevent sampling of conformations.
Hence, kinetics would not converge towards the Native structure but would diffuse
freely among the extended structures subset.

At $B_s = -7.0$, the couplings are always attractive at low temperature.
The cold denaturation disappears and
for $B_s < -7.0$, only the warm transition remains. 
A critical point occurs for $B_s^{(c)} = -5.75$ and $T^{(c)} = 0.53$.
At this point,  one has $\tijm \approx \tijp$.

\section{Protein Kinetics}
The folding time, $\tf$, only defined in the native phase, 
is the mean Monte Carlo steps needed to reach Nat for the first time,
averaged over 1000 trajectories\cite{Sali1994a} . 
Each trajectory 
starts with a random conformation and Monte Carlo simulations
using the corner flip, the tail and the crankshaft moves used in 
\cite{Chan1994,Collet2003b}, are performed.
The folding times are plotted versus $B_s$ and $T$ in fig.\ref{tfold}.

\begin{figure}[htbp]
\centerline{\includegraphics[width=3.0in]{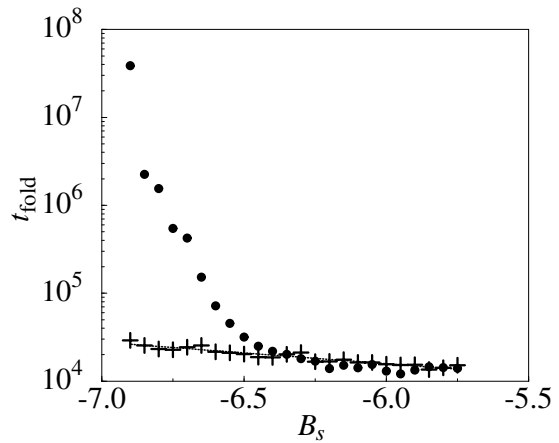}}
%\vskip0.2cm
\caption{\label{tfTcTw}.
Folding times at the temperatures of denaturation as a function of the solvent quality.
The filled circle are for the temperature of cold denaturation and the + are for the
warm denaturation.}
\end{figure}

\begin{figure}[hbtp]
\centerline{\includegraphics[width=3.0in]{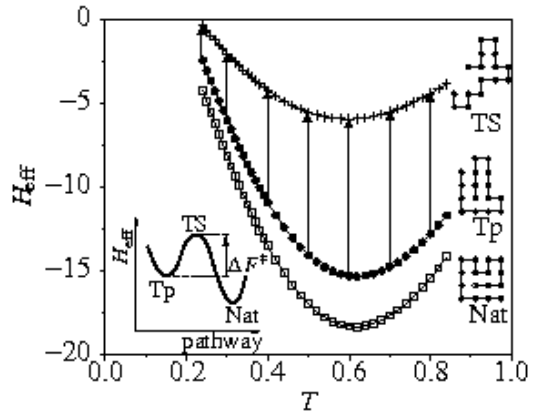}}
\caption{\label{ftot}
Inset: schematic free energy landscape of a folding pathway 
from the trap (Tp) to the native (Nat) structure through the
transition state (TS) with the activation barrier (arrows)
$\Delta F^\ddagger(T) = {\cal H}_{\rm eff}^{(\rm TS)} 
	- {\cal H}_{\rm eff}^{(\rm Trap)}$.
Main: free energy of the native (square), the trap (filled circle) 
conformations and the transition state (+)
versus temperature for $B_s = -6.25$. The arrows show some 
free energy barriers $\Delta F^\ddagger(T)$.
}
\end{figure}

For $-5.75 > B_s > -6.4$, one has 
$\tf(T_c) \approx \tf(T_w)$.
That is to say, the folding times are always the same at the temperatures
of denaturation in this case (see fig.\ref{tfTcTw}).
%\vbox{
%}
This property may be understood as follow. 
By definition, the equilibrium probability of occurrence
of the native structure equals 1/2 at $T_c$ and $T_w$.
Moreover, it may be seen on the example of fig.\ref{ftot},
that, for $Bs=-6.25$, the temperatures of transition are 
$T_w=0.83$ and $T_c=0.24$ and the effective Hamiltonian of the native structure
for these temperatures are 
${\cal H}_{\rm eff}^{\rm Nat}(T_c) = -4.2$ and
${\cal H}_{\rm eff}^{\rm Nat}(T_w) = -14.2$ leading to 
$${\cal H}_{\rm eff}^{\rm Nat}(T_c)/T_c \approx
{\cal H}_{\rm eff}^{\rm Nat}(T_w) / T_w$$
As, 
$$\frac{\exp[-{\cal H}_{\rm eff}^{\rm Nat}(T_c)/T_c ] }{Z(T_c)}  =
  \frac{\exp[-{\cal H}_{\rm eff}^{\rm Nat}(T_w)/T_w ] }{Z(T_w)} $$
one deduces the equality of the partition functions, $Z(T_c) = Z(T_w)$.
Thus, the probability of occurrence of each extended structure,
having a zero effective Hamiltonian, is simply one over the partition function
and is the same at $T_c$ and $T_w$.
Rationally, to satisfy to the equality of the partition functions, one supposes that
the conformational effective Hamiltonians satisfy to 
$ {\cal H}_{\rm eff}^{(m)}(T_c)/T_c \approx
{\cal H}_{\rm eff}^{(m)}(T_w) / T_w$ whatever the structure $m$.
The configurational spaces where the effective Hamiltonians are roughly linearly scaled by 
the temperatures may thus be expected for these temperatures.
This result holds whatever the solvent quality.
To be more precise, the equilibrium probabilities of each
structure remains almost constant at the temperatures of denaturation
when the solvent quality is varied.
As a consequence, the transition rates between two conformations $m$ and $n$,
$w(m \rightarrow n, T) = \exp[(\Hn(T) - \H(T))/2T]$,
are almost constant at the denaturations temperatures for all $B_s>-6.4$,
leading to this (quasi) equality of the folding times at $T_c$ and $T_w$ 

For $B_s \approx -7.0$, the temperature of cold denaturation tends toward 0.
The conformational landscape still exhibits a well pronounced 
effective Hamiltonian minimum for the native structure, but as the temperature
becomes very small the transition rates becomes either quasi null or huge.
It takes huge amount of time to overcome some local energetic barriers.
Thus the folding time tends to infinity at $T_c$ 
because the kinetics is then frozen.

Then, folding time at $T_c$ increases rapidly as 
$B_s$ decreases from -6.4 to -7.0.

\vskip0.3cm

For each value of $B_s$, the folding time goes 
through a maximum value at a temperature noted  $T^*(B_s)$.
The region of the curves where $T>T^*$, simply confirms
that the time needed for a random conformation to reach the 
native structure increases as the temperature is decreased.
However, the behavior of the curves for $T<T^*$ is less usual.
It corresponds to an increasing of the speed of the kinetics with respect
to a decrease of the temperature. 
This result is not in accordance with standard Monte
Carlo simulations and its explanation is given below.

For each temperature, the kinetic trap\cite{Cieplak1998} (shown in fig.\ref{ftot}) is found using 
Monte Carlo algorithm by noting
the most occurrent conformation during the 1000 trajectories performed
to calculated the folding time.
%\vbox{
%}
It is observed that the trap do not depend on the temperature.
The trap present a great similarity with the native structure.
Six contacts among the nine contacts of Nat are observed in trap.
Thus,  its effective Hamiltonian is very close from that
of Nat showing that this structure could be a good candidate for the trap
of the system.
Second, there is no physical pathway from trap to Nat keeping unbroken the whole
set of contacts.
That is to say, in a typical trajectory from the trap to Nat, each contact of
the trap has to be broken and some of them created again 
in an other global arrangement.
Obviously, they are not simultaneously broken. If it were the case, 
the trajectories would contain some extended structures without contacts
which would have a very great difference of effective Hamiltonian with the trap.
Consequently, even step by step, the transition rate for such sub-pathway would be quasi-null.
Thus, in more favorable pathways, some contacts are broken while some others are created
and the chain walks through local rearrangements toward Nat without reaching
any extended conformation.
A particular structure of the sequence is the transition state.
It is determined by performing a new simulation of 1000 trajectories where
the trap is always the first conformation.
For each trajectory, the conformation of highest value of the effective Hamiltonian
is considered as a possible transition state.
The transition state (TS) is the structure with the lowest value of effective 
Hamiltonian among the sampling of possible transition states 
collected over the 1000 trajectories (see inset of fig.\ref{ftot}).
Among all the possible pathways from trap to Nat, those passing through
TS are the less energetically costly.
One must note that this is not the usual definition of the transition state adopted in
the theory of protein folding where the TS is not a unique structure but an ensemble of 
configurations of highest free energy along the path or paths between unfolded macrostates
and the native structure \cite{Onuchic1996}. Here, the defintion of the TS is that used
in the theory of the simple gaz chemical reactions.
Moreover, as explained above, it can be seen that TS has a weak similarity 
with Nat or with the trap.

The key role of the trap in the folding process is explained as follow.
Some trajectories, starting by a random conformations, fall in the trap valley.
To leave this structure the chain has to overcome the largest barrier of the
system. On the other hand, the trajectories which do not reach the trap valley, 
walk down the native structure, passing quickly over smaller barriers.
Thus, only the trap activation barrier is considered in the folding analysis.

\begin{figure}[hbtp]
\vskip0.8cm
\centerline{\includegraphics[width=2.9in]{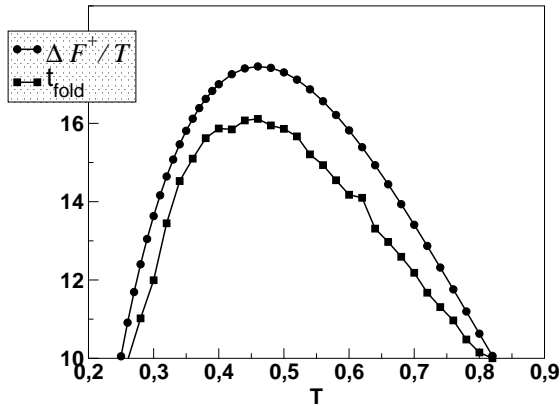}}
\caption{\label{arrh}
Logarithm of the folding time (filled square) and 
$\Delta F^\ddagger(T) / T$ (filled circle) versus the temperature
for $B_s = -6.25$.
}
\end{figure}

For $B_s=-6.25$, the values of the effective Hamiltonian for Nat, the trap and TS 
are reported on fig.\ref{ftot}.
The activation barrier between the trap and Nat structures, $\Delta F^\ddagger= {\cal H}_{\rm eff}^{(\rm TS)} 
	- {\cal H}_{\rm eff}^{(\rm Trap)}$, shown on fig.\ref{ftot}, 
obviously depends on the temperature. For $T> 0.60$, $\Delta F^\ddagger$ is almost constant whereas it decreases very quickly with the temperature 
for $T<0.45$.
Moreover, the kinetic theory indicates that 
the time needed to escape from the trap valley is in proportion 
to $\exp(\Delta F^\ddagger / T)$.
Figure \ref{arrh} showed the very sharp decreasing of
$\Delta F^\ddagger / T$ with respect to a decrease of the temperature for 
$T> 0.45$. One must noted the quasi equality of this last quantity
at the temperatures of denaturation.
%\vbox{
%}

The values of $\Delta F^\ddagger /T$ and that of the folding time
starting from a random structure, reported in fig.\ref{arrh}, 
 present maxima at the temperature, $T_1 \approx 0.45$.
For $T < T_1$, the rate $\Delta F^\ddagger /T$ decreases as the
temperature is decreased because $\Delta F^\ddagger$ decreases faster than $T$.
For $T > T_1$,  the rate $\Delta F^\ddagger /T$ decreases as the
temperature is increased because $\Delta F^\ddagger$ is almost constant
and, obviously, $1/T$ decreases. 
As, the folding time is in proportion to $\exp(\Delta F^\ddagger (T)/T)$, this result
permits to elucidate why 
the lower the temperature, the faster is the kinetics at low temperature and 
the higher the temperature, the faster also is the kinetics at high temperature.

\section{Conclusion} 
Our model predicts and explains unusual behavior
of the folding times due to the flatness of the conformation
space as the temperature decreases towards $T_c$.
In the native region, far from the denaturation temperatures,
 $P_{\rm eq}^{\rm Nat}$ tends to 1, leading to 
a  very rough conformational landscape.
On the opposite, around $T_c$, effective Hamiltonian of all conformations are 
quasi equal and the free energy landscape is quasi flat.
For low temperatures, the difference of effective Hamiltonian between two
conformations decreases faster than the temperature.
This leveling of the conformational space close to the cold transition
leads to a drastic decrease of the free energy barriers of the configurational 
space leading to an acceleration of the kinetics as the temperature decreases.
Last and unfortunately, at the present time, they are no experimental results 
available for the folding kinetics of proteins
at temperatures close to that of cold denaturation.
which should be compared to this theoretical prediction.

\newpage
%\listoffigures

\begin{thebibliography}{36}
\expandafter\ifx\csname natexlab\endcsname\relax\def\natexlab#1{#1}\fi
\expandafter\ifx\csname bibnamefont\endcsname\relax
  \def\bibnamefont#1{#1}\fi
\expandafter\ifx\csname bibfnamefont\endcsname\relax
  \def\bibfnamefont#1{#1}\fi
\expandafter\ifx\csname citenamefont\endcsname\relax
  \def\citenamefont#1{#1}\fi
\expandafter\ifx\csname url\endcsname\relax
  \def\url#1{\texttt{#1}}\fi
\expandafter\ifx\csname urlprefix\endcsname\relax\def\urlprefix{URL }\fi
\providecommand{\bibinfo}[2]{#2}
\providecommand{\eprint}[2][]{\url{#2}}

\bibitem[{\citenamefont{Anfinsen et~al.}(1961)\citenamefont{Anfinsen, Haber,
  Sela, and F.H.}}]{Anfinsen1961}
\bibinfo{author}{\bibfnamefont{C.~B.} \bibnamefont{Anfinsen}},
  \bibinfo{author}{\bibfnamefont{E.}~\bibnamefont{Haber}},
  \bibinfo{author}{\bibfnamefont{M.}~\bibnamefont{Sela}}, \bibnamefont{and}
  \bibinfo{author}{\bibfnamefont{W.}~\bibnamefont{F.H.}},
  \bibinfo{journal}{Proc. Natl. Acad. Sci.} \textbf{\bibinfo{volume}{47}},
  \bibinfo{pages}{1309} (\bibinfo{year}{1961}).

\bibitem[{\citenamefont{Bryngelson and Wolynes}(1987)}]{Bryngelson1987}
\bibinfo{author}{\bibfnamefont{J.~D.} \bibnamefont{Bryngelson}}
  \bibnamefont{and} \bibinfo{author}{\bibfnamefont{P.~G.}
  \bibnamefont{Wolynes}}, \bibinfo{journal}{Proc. Natl. Acad. Sci. USA}
  \textbf{\bibinfo{volume}{84}}, \bibinfo{pages}{7524} (\bibinfo{year}{1987}).

\bibitem[{\citenamefont{Shakhnovich and Gutin}(1990)}]{Shakhnovich1990a}
\bibinfo{author}{\bibfnamefont{E.~I.} \bibnamefont{Shakhnovich}}
  \bibnamefont{and} \bibinfo{author}{\bibfnamefont{A.~M.} \bibnamefont{Gutin}},
  \bibinfo{journal}{Nature} \textbf{\bibinfo{volume}{346}},
  \bibinfo{pages}{773} (\bibinfo{year}{1990}).

\bibitem[{\citenamefont{Dinner et~al.}(1994)\citenamefont{Dinner, \v{S}ali,
  Karplus, and Shakhnovich}}]{Dinner1994}
\bibinfo{author}{\bibfnamefont{A.}~\bibnamefont{Dinner}},
  \bibinfo{author}{\bibfnamefont{A.}~\bibnamefont{\v{S}ali}},
  \bibinfo{author}{\bibfnamefont{M.}~\bibnamefont{Karplus}}, \bibnamefont{and}
  \bibinfo{author}{\bibfnamefont{E.}~\bibnamefont{Shakhnovich}},
  \bibinfo{journal}{J. Chem. Phys.} \textbf{\bibinfo{volume}{101}},
  \bibinfo{pages}{1444} (\bibinfo{year}{1994}).

\bibitem[{\citenamefont{Shakhnovich}(1994)}]{Shakhnovich1994}
\bibinfo{author}{\bibfnamefont{E.~I.} \bibnamefont{Shakhnovich}},
  \bibinfo{journal}{Phys. Rev. Lett.} \textbf{\bibinfo{volume}{72}},
  \bibinfo{pages}{3907} (\bibinfo{year}{1994}).

\bibitem[{\citenamefont{Gutin et~al.}(1995)\citenamefont{Gutin, Abkevich, and
  Shakhnovich}}]{Gutin1995b}
\bibinfo{author}{\bibfnamefont{A.~M.} \bibnamefont{Gutin}},
  \bibinfo{author}{\bibfnamefont{V.~I.} \bibnamefont{Abkevich}},
  \bibnamefont{and} \bibinfo{author}{\bibfnamefont{E.~I.}
  \bibnamefont{Shakhnovich}}, \bibinfo{journal}{Proc. Natl. Acad. Sci. USA}
  \textbf{\bibinfo{volume}{92}}, \bibinfo{pages}{1282} (\bibinfo{year}{1995}).

\bibitem[{\citenamefont{Kumar et~al.}(2006)\citenamefont{Kumar, Prabhu, Rao,
  and Bhuyan}}]{Kumar2006}
\bibinfo{author}{\bibfnamefont{R.}~\bibnamefont{Kumar}},
  \bibinfo{author}{\bibfnamefont{A.}~\bibnamefont{Prabhu}},
  \bibinfo{author}{\bibfnamefont{D.}~\bibnamefont{Rao}}, \bibnamefont{and}
  \bibinfo{author}{\bibfnamefont{A.}~\bibnamefont{Bhuyan}},
  \bibinfo{journal}{J. Mol. Biol.} \textbf{\bibinfo{volume}{364}},
  \bibinfo{pages}{483} (\bibinfo{year}{2006}).

\bibitem[{\citenamefont{Pastore et~al.}(2007)\citenamefont{Pastore, Martin,
  Politou, Kondapalli, and Temussi}}]{Pastore2007}
\bibinfo{author}{\bibfnamefont{A.}~\bibnamefont{Pastore}},
  \bibinfo{author}{\bibfnamefont{S.}~\bibnamefont{Martin}},
  \bibinfo{author}{\bibfnamefont{A.}~\bibnamefont{Politou}},
  \bibinfo{author}{\bibfnamefont{T.}~\bibnamefont{Kondapalli},
  \bibfnamefont{K.C.~Stemmler}}, \bibnamefont{and}
  \bibinfo{author}{\bibfnamefont{P.}~\bibnamefont{Temussi}},
  \bibinfo{journal}{J. Am. Chem. Soc.} \textbf{\bibinfo{volume}{129}},
  \bibinfo{pages}{5374} (\bibinfo{year}{2007}).

\bibitem[{\citenamefont{Hadi-Alijanvand
  et~al.}(2007)\citenamefont{Hadi-Alijanvand, Ahmad, and
  Moosavi-Movahedi}}]{Hadi2007}
\bibinfo{author}{\bibfnamefont{H.}~\bibnamefont{Hadi-Alijanvand}},
  \bibinfo{author}{\bibfnamefont{F.}~\bibnamefont{Ahmad}}, \bibnamefont{and}
  \bibinfo{author}{\bibfnamefont{A.}~\bibnamefont{Moosavi-Movahedi}},
  \bibinfo{journal}{Protein Journal} \textbf{\bibinfo{volume}{26}},
  \bibinfo{pages}{395} (\bibinfo{year}{2007}).

\bibitem[{\citenamefont{Whitten et~al.}(2007)\citenamefont{Whitten, Kurtz,
  Pometun, Wand, and Hilsen}}]{Whitten2007}
\bibinfo{author}{\bibfnamefont{S.}~\bibnamefont{Whitten}},
  \bibinfo{author}{\bibfnamefont{A.}~\bibnamefont{Kurtz}},
  \bibinfo{author}{\bibfnamefont{M.}~\bibnamefont{Pometun}},
  \bibinfo{author}{\bibfnamefont{A.}~\bibnamefont{Wand}}, \bibnamefont{and}
  \bibinfo{author}{\bibfnamefont{V.}~\bibnamefont{Hilser}},
  \bibinfo{journal}{Biochemistry} \textbf{\bibinfo{volume}{45}},
  \bibinfo{pages}{10163} (\bibinfo{year}{2007}).

\bibitem[{\citenamefont{Privalov and Makhatadze}(1989)}]{Privalov1989}
\bibinfo{author}{\bibfnamefont{P.~L.} \bibnamefont{Privalov}} \bibnamefont{and}
  \bibinfo{author}{\bibfnamefont{G.~I.} \bibnamefont{Makhatadze}},
  \bibinfo{journal}{J. Mol. Biol.} \textbf{\bibinfo{volume}{205}},
  \bibinfo{pages}{737} (\bibinfo{year}{1989}).

\bibitem[{\citenamefont{Kauzmann}(1959)}]{Kauzmann1959}
\bibinfo{author}{\bibfnamefont{W.}~\bibnamefont{Kauzmann}},
  \bibinfo{journal}{Adv. Protein Chem.} \textbf{\bibinfo{volume}{14}},
  \bibinfo{pages}{1} (\bibinfo{year}{1959}).

\bibitem[{\citenamefont{Warshel and Lifson}(1970)}]{Warshel1970}
\bibinfo{author}{\bibfnamefont{A.}~\bibnamefont{Warshel}} \bibnamefont{and}
  \bibinfo{author}{\bibfnamefont{S.}~\bibnamefont{Lifson}},
  \bibinfo{journal}{J. Chem. Phys.} \textbf{\bibinfo{volume}{53}},
  \bibinfo{pages}{582} (\bibinfo{year}{1970}).

\bibitem[{\citenamefont{Dill}(1990)}]{Dill1990}
\bibinfo{author}{\bibfnamefont{K.}~\bibnamefont{Dill}},
  \bibinfo{journal}{Biochemistry} \textbf{\bibinfo{volume}{29}},
  \bibinfo{pages}{7133} (\bibinfo{year}{1990}).

\bibitem[{\citenamefont{Premilat and Collet}(1997)}]{Premilat1997}
\bibinfo{author}{\bibfnamefont{S.}~\bibnamefont{Premilat}} \bibnamefont{and}
  \bibinfo{author}{\bibfnamefont{O.}~\bibnamefont{Collet}},
  \bibinfo{journal}{Europhysics Letters} \textbf{\bibinfo{volume}{39}},
  \bibinfo{pages}{575} (\bibinfo{year}{1997}).

\bibitem[{\citenamefont{Collet and Premilat}(1996)}]{Collet1996}
\bibinfo{author}{\bibfnamefont{O.}~\bibnamefont{Collet}} \bibnamefont{and}
  \bibinfo{author}{\bibfnamefont{S.}~\bibnamefont{Premilat}},
  \bibinfo{journal}{Journal of Molecular Structure. Theochem}
  \textbf{\bibinfo{volume}{363}}, \bibinfo{pages}{151} (\bibinfo{year}{1996}).

\bibitem[{\citenamefont{Frauenfelder et~al.}(2006)\citenamefont{Frauenfelder,
  Fenimore, Chen, and McMahon}}]{Frauenfelder2006}
\bibinfo{author}{\bibfnamefont{H.}~\bibnamefont{Frauenfelder}},
  \bibinfo{author}{\bibfnamefont{P.}~\bibnamefont{Fenimore}},
  \bibinfo{author}{\bibfnamefont{G.}~\bibnamefont{Chen}}, \bibnamefont{and}
  \bibinfo{author}{\bibfnamefont{B.}~\bibnamefont{McMahon}},
  \bibinfo{journal}{Proc., Natl., Acad., Sci., USA}
  \textbf{\bibinfo{volume}{103}}, \bibinfo{pages}{15469}
  (\bibinfo{year}{2006}).

\bibitem[{\citenamefont{Collet}(2001)}]{Collet2001}
\bibinfo{author}{\bibfnamefont{O.}~\bibnamefont{Collet}},
  \bibinfo{journal}{Europhys. Letters} \textbf{\bibinfo{volume}{53}},
  \bibinfo{pages}{93} (\bibinfo{year}{2001}).

\bibitem[{\citenamefont{Collet}(2005)}]{Collet2005}
\bibinfo{author}{\bibfnamefont{O.}~\bibnamefont{Collet}},
  \bibinfo{journal}{Europhys. Letters} \textbf{\bibinfo{volume}{72}},
  \bibinfo{pages}{301} (\bibinfo{year}{2005}).

\bibitem[{\citenamefont{Silverstein et~al.}(1999)\citenamefont{Silverstein,
  Haymet, and Dill}}]{Silverstein1999}
\bibinfo{author}{\bibfnamefont{K.~A.~T.} \bibnamefont{Silverstein}},
  \bibinfo{author}{\bibfnamefont{A.~D.~J.} \bibnamefont{Haymet}},
  \bibnamefont{and} \bibinfo{author}{\bibfnamefont{K.~A.} \bibnamefont{Dill}},
  \bibinfo{journal}{J. Chem. Phys.} \textbf{\bibinfo{volume}{111}},
  \bibinfo{pages}{8000} (\bibinfo{year}{1999}).

\bibitem[{\citenamefont{Hansen et~al.}(1999)\citenamefont{Hansen, Jensen,
  Sneppen, and Zocchi}}]{Hansen1999}
\bibinfo{author}{\bibfnamefont{A.}~\bibnamefont{Hansen}},
  \bibinfo{author}{\bibfnamefont{M.~H.} \bibnamefont{Jensen}},
  \bibinfo{author}{\bibfnamefont{K.}~\bibnamefont{Sneppen}}, \bibnamefont{and}
  \bibinfo{author}{\bibfnamefont{G.}~\bibnamefont{Zocchi}},
  \bibinfo{journal}{cond-mat/9905357}  (\bibinfo{year}{1999}).

\bibitem[{\citenamefont{De~Los~Rios and Caldarelli}(2000)}]{DeLosRios2000}
\bibinfo{author}{\bibfnamefont{P.}~\bibnamefont{De~Los~Rios}} \bibnamefont{and}
  \bibinfo{author}{\bibfnamefont{G.}~\bibnamefont{Caldarelli}},
  \bibinfo{journal}{Phys. Rev. E} \textbf{\bibinfo{volume}{62}},
  \bibinfo{pages}{8449} (\bibinfo{year}{2000}).

\bibitem[{\citenamefont{Roccatano et~al.}(2004)\citenamefont{Roccatano,
  Di~Nola, and Amadei}}]{Roccatano2004}
\bibinfo{author}{\bibfnamefont{D.}~\bibnamefont{Roccatano}},
  \bibinfo{author}{\bibfnamefont{A.}~\bibnamefont{Di~Nola}}, \bibnamefont{and}
  \bibinfo{author}{\bibfnamefont{A.}~\bibnamefont{Amadei}},
  \bibinfo{journal}{J. Phys. Chem. B.} \textbf{\bibinfo{volume}{108(18)}},
  \bibinfo{pages}{5756} (\bibinfo{year}{2004}).

\bibitem[{\citenamefont{Lopez et~al.}(2008)\citenamefont{Lopez, Darst, and
  Rossky}}]{Lopez2008}
\bibinfo{author}{\bibfnamefont{C.~F.} \bibnamefont{Lopez}},
  \bibinfo{author}{\bibfnamefont{R.~K.} \bibnamefont{Darst}}, \bibnamefont{and}
  \bibinfo{author}{\bibfnamefont{P.~J.} \bibnamefont{Rossky}},
  \bibinfo{journal}{J. Phys. Chem. B.} \textbf{\bibinfo{volume}{112(19)}},
  \bibinfo{pages}{5961} (\bibinfo{year}{2008}).

\bibitem[{\citenamefont{Patel et~al.}(2008)\citenamefont{Patel, Debenedetti,
  Stillinger, and Rossky}}]{Patel2008}
\bibinfo{author}{\bibfnamefont{B.~A.} \bibnamefont{Patel}},
  \bibinfo{author}{\bibfnamefont{P.~G.} \bibnamefont{Debenedetti}},
  \bibinfo{author}{\bibfnamefont{F.~H.} \bibnamefont{Stillinger}},
  \bibnamefont{and} \bibinfo{author}{\bibfnamefont{P.~J.}
  \bibnamefont{Rossky}}, \bibinfo{journal}{J. Chem. Phys.}
  \textbf{\bibinfo{volume}{128}}, \bibinfo{pages}{175102}
  (\bibinfo{year}{2008}).

\bibitem[{\citenamefont{Dias et~al.}(2008)\citenamefont{Dias, Ala-Nissila,
  Karttunen, Vattulainen, and Grant}}]{Dias2008}
\bibinfo{author}{\bibfnamefont{C.~L.} \bibnamefont{Dias}},
  \bibinfo{author}{\bibfnamefont{T.}~\bibnamefont{Ala-Nissila}},
  \bibinfo{author}{\bibfnamefont{M.}~\bibnamefont{Karttunen}},
  \bibinfo{author}{\bibfnamefont{I.}~\bibnamefont{Vattulainen}},
  \bibnamefont{and} \bibinfo{author}{\bibfnamefont{M.}~\bibnamefont{Grant}},
  \bibinfo{journal}{Phys. Rev. Lett.} \textbf{\bibinfo{volume}{100}},
  \bibinfo{pages}{118101} (\bibinfo{year}{2008}).

\bibitem[{\citenamefont{Metro\-polis et~al.}(1953)\citenamefont{Metro\-polis,
  Rosenbluth, Rosenbluth, Teller, and Teller}}]{Metropolis1953}
\bibinfo{author}{\bibfnamefont{N.}~\bibnamefont{Metro\-polis}},
  \bibinfo{author}{\bibfnamefont{A.~W.} \bibnamefont{Rosenbluth}},
  \bibinfo{author}{\bibfnamefont{M.~N.} \bibnamefont{Rosenbluth}},
  \bibinfo{author}{\bibfnamefont{A.~H.} \bibnamefont{Teller}},
  \bibnamefont{and} \bibinfo{author}{\bibfnamefont{E.}~\bibnamefont{Teller}},
  \bibinfo{journal}{J. Chem. Phys.} \textbf{\bibinfo{volume}{21}},
  \bibinfo{pages}{1087} (\bibinfo{year}{1953}).

\bibitem[{\citenamefont{Chan and Dill}(1994)}]{Chan1994}
\bibinfo{author}{\bibfnamefont{H.~S.} \bibnamefont{Chan}} \bibnamefont{and}
  \bibinfo{author}{\bibfnamefont{K.~A.} \bibnamefont{Dill}},
  \bibinfo{journal}{J. Chem. Phys.} \textbf{\bibinfo{volume}{100}},
  \bibinfo{pages}{9238} (\bibinfo{year}{1994}).

\bibitem[{\citenamefont{Ben-Na\"{i}m}(1970)}]{Bennaim1970}
\bibinfo{author}{\bibfnamefont{A.}~\bibnamefont{Ben-Na\"{i}m}},
  \bibinfo{journal}{J. Chem. Phys.} \textbf{\bibinfo{volume}{54}},
  \bibinfo{pages}{3682} (\bibinfo{year}{1970}).

\bibitem[{\citenamefont{Muller}(1990)}]{Muller1990}
\bibinfo{author}{\bibfnamefont{N.}~\bibnamefont{Muller}},
  \bibinfo{journal}{Acc. Chem. Res.} \textbf{\bibinfo{volume}{23}},
  \bibinfo{pages}{23} (\bibinfo{year}{1990}).

\bibitem[{\citenamefont{Lee and Graziano}(1996)}]{Lee1996}
\bibinfo{author}{\bibfnamefont{B.}~\bibnamefont{Lee}} \bibnamefont{and}
  \bibinfo{author}{\bibfnamefont{G.}~\bibnamefont{Graziano}},
  \bibinfo{journal}{J. Am. Chem. Soc.} \textbf{\bibinfo{volume}{118}},
  \bibinfo{pages}{5163} (\bibinfo{year}{1996}).

\bibitem[{\citenamefont{Baldwin}(1986)}]{Baldwin1986}
\bibinfo{author}{\bibfnamefont{R.}~\bibnamefont{Baldwin}},
  \bibinfo{journal}{Proc., Natl., Acad., Sci., USA}
  \textbf{\bibinfo{volume}{83}}, \bibinfo{pages}{8069} (\bibinfo{year}{1986}).

\bibitem[{\citenamefont{{\v{S}}ali et~al.}(1994)\citenamefont{{\v{S}}ali,
  Shakhnovich, and Karplus}}]{Sali1994a}
\bibinfo{author}{\bibfnamefont{A.}~\bibnamefont{{\v{S}}ali}},
  \bibinfo{author}{\bibfnamefont{E.}~\bibnamefont{Shakhnovich}},
  \bibnamefont{and} \bibinfo{author}{\bibfnamefont{M.}~\bibnamefont{Karplus}},
  \bibinfo{journal}{J. Mol. Biol.} \textbf{\bibinfo{volume}{235}},
  \bibinfo{pages}{1614} (\bibinfo{year}{1994}).

\bibitem[{\citenamefont{Collet}(2003)}]{Collet2003b}
\bibinfo{author}{\bibfnamefont{O.}~\bibnamefont{Collet}},
  \bibinfo{journal}{Phys. Rev. E} \textbf{\bibinfo{volume}{67}},
  \bibinfo{pages}{061912} (\bibinfo{year}{2003}).

\bibitem[{\citenamefont{Cieplak et~al.}(1998)\citenamefont{Cieplak, Henkel,
  Karbowski, and Banavar}}]{Cieplak1998}
\bibinfo{author}{\bibfnamefont{M.}~\bibnamefont{Cieplak}},
  \bibinfo{author}{\bibfnamefont{M.}~\bibnamefont{Henkel}},
  \bibinfo{author}{\bibfnamefont{J.}~\bibnamefont{Karbowski}},
  \bibnamefont{and} \bibinfo{author}{\bibfnamefont{J.}~\bibnamefont{Banavar}},
  \bibinfo{journal}{Phys. Rev. Lett.} \textbf{\bibinfo{volume}{80}},
  \bibinfo{pages}{3654} (\bibinfo{year}{1998}).

\bibitem[{\citenamefont{Onuchic et~al.}(1996)\citenamefont{Onuchic, Socci,
  Luthey-Schulten, and Wolynes}}]{Onuchic1996}
\bibinfo{author}{\bibfnamefont{J.}~\bibnamefont{Onuchic}},
  \bibinfo{author}{\bibfnamefont{D.}~\bibnamefont{Socci}},
  \bibinfo{author}{\bibfnamefont{Z.}~\bibnamefont{Luthey-Schulten}},
  \bibnamefont{and} \bibinfo{author}{\bibfnamefont{P.}~\bibnamefont{Wolynes}},
  \bibinfo{journal}{Folding \& Design} \textbf{\bibinfo{volume}{1}},
  \bibinfo{pages}{441} (\bibinfo{year}{1996}).

\end{thebibliography}
\end{document}